\documentclass[%
 reprint,
superscriptaddress,
nofootinbib,
 amsmath,amssymb,
 aps, braket,
pra
]{revtex4-1}
\usepackage{pbox}
\usepackage{epstopdf}
\usepackage{tikz}
\usepackage{bbm}
\usetikzlibrary{quantikz}
\usepackage{amsmath}
\usepackage{mathpazo}
\usepackage{graphicx}% Include figure files
\usepackage{dcolumn}% Align table columns on decimal point
\usepackage{bm}% bold math
\usepackage{physics}
\usepackage{multirow}
\usepackage{subfiles}
\usepackage[hidelinks]{hyperref}
\usepackage{amsthm}
\usepackage{apptools}
\usepackage{chngcntr}
\usepackage{mathtools}
\usepackage{subcaption}
\usepackage{color}
\usepackage{transparent}
\usepackage{mathrsfs}
\usepackage{algorithm}
\usepackage{algpseudocode}
\usepackage{ulem}
\AtAppendix{\counterwithin{theorem}{section}}
\usepackage{tikz, tikz-cd}

\DeclareMathOperator*{\argmin}{arg\,min}
\newtheorem{theorem}{Theorem}
\newtheorem{Theorem}[theorem]{Theorem}
\newtheorem{Definition}{Definition}

\newtheorem{Question}[theorem]{Question}
\newtheorem{Lemma}[theorem]{Lemma}

\newtheorem{Example}{Example}

\newcommand{\data}{\Omega}

\newcommand{\stabgrp}{\mathcal{S}}
\newcommand{\cnot}{\mathsf{CNOT}}
\newcommand{\Pauligrp}{\mathcal{P}}
\newcommand{\code}{\mathcal{C}}

\newcommand{\Hilbert}{\mathcal{H}}

\newcommand{\gadget}{\mathcal{G}}

\newcommand{\ztwo}{\mathbb{F}_2}

\newcommand{\im}{\mathrm{im\;}}
\newcommand{\transpose}{\mathsf{T}}

\newcommand{\dataerr}{\mathcal{D}}
\newcommand{\measerr}{\mathcal{M}}
\newcommand{\decoder}{\mathsf{Dec}}
\newcommand{\mwedecoder}{\mathsf{Dec}_{\textrm{MWE}}}
\newcommand{\mwpmdecoder}{\mathsf{Dec}_{\textrm{MWPM}}}
\newcommand{\network}{U}

\makeatletter
\def\blfootnote{\xdef\@thefnmark{}\@footnotetext}
\makeatother
\makeatletter
\newenvironment{algbox}[1][htb]{%
    \renewcommand{\ALG@name}{Procedure}% Update algorithm name
   \begin{algorithm}[#1]%
  }{\end{algorithm}}
\makeatother

\begin{document}

\preprint{APS/123-QED}

\title{
Between Shor and Steane: A unifying construction for measuring error syndromes
}% Force line breaks with \\

\author{Shilin Huang}
\email{shilin.huang@duke.edu}
\affiliation{Department of Electrical and Computer Engineering, Duke University, Durham, NC 27708, USA}

\author{Kenneth R. Brown}
\email{kenneth.r.brown@duke.edu}
\affiliation{Department of Electrical and Computer Engineering, Duke University, Durham, NC 27708, USA}
 \affiliation{Department of Physics, Duke University, Durham, NC 27708, USA}

\affiliation{Department of Chemistry, Duke University, Durham, NC 27708, USA}

\date{\today}% It is always \today, today,
\begin{abstract}
Fault-tolerant quantum error correction requires the measurement of error syndromes in a way that minimizes correlated errors on the quantum data.  Steane and Shor ancilla are two well-known methods for fault-tolerant syndrome extraction. In this paper, we  find a unifying construction that generates a family of ancilla blocks that  interpolate between Shor and Steane. This family increases the complexity of ancilla construction in exchange for reducing the rounds of measurement required to fault-tolerantly measure the error. 
We then apply this construction to  the toric code of size $L\times L$ and find that blocks of size $m\times m$ can be used to decode errors in $O(L/m)$ rounds of measurements. Our method can be applied to any Calderbank-Shor-Steane codes and presents a new direction for optimizing fault-tolerant quantum computation. 
\end{abstract}

\maketitle

\section{Introduction}

The theory of quantum error correction~\cite{Shor:1995b, Aharonov:1997, Knill1997, kitaev2003fault, calderbank1996good,Steane:1996,calderbank1997orthgonal,calderbank1998gf4, shor1996fault, divincenzo1996fault} opens the path towards large-scale quantum computation. Stabilizer codes are a conventional choice of quantum error-correcting codes~\cite{calderbank1997orthgonal,calderbank1998gf4,Gottesman:1997}. Error corrections is performed conditional on the measurement outcomes of a set of code stabilizers, also known as the error syndrome bit string.
The syndrome extraction circuits need to be fault-tolerant~\cite{shor1996fault,Aliferis:2006,steane1997active,Steane:2003,Knill2005nature, Chao2018, Chao2018few, chao2020flag} as the act of measuring sydromes also introduces extra errors on the quantum systems. 

The first fault-tolerant syndrome extraction scheme was proposed by Shor~\cite{shor1996fault,divincenzo1996fault}. 
In Shor's scheme, each syndrome bit is extracted from the data qubits to a
verified ancilla cat state by transversal two-qubit gates. 
As transversal operations limit the error propagation, no high-weight correlated errors can occur on the data qubits if the cat states are verified by post-selection.
The value of the syndrome bit is the parity of the transversal measurement outcome of the cat state. 
As any measurement error will flip the syndrome bit, for a stabilizer code of distance $d$,
one needs to repeat the syndrome extraction for $O(d^2)$ rounds to guarantee fault-tolerance. 
Utilizing the information of the code structure,
the time overhead can be significantly reduced on particular codes~\cite{Dennis:2002,hector2015single,Fawzi2018constant,Campbell2019,delfosse2020beyond,delfosse2020short}.

Optimizing Shor's scheme is an active area of research building off substantial progress since its invention. 
For example, for low-weight stabilizers, 
ancilla post-selection could be avoided by decoding the ancilla cat states~\cite{divincenzo2007effective,Yoder:2017b,Chao2018}. 
On specific stabilizer codes, the space overhead can be reduced by allowing non-transversal data-ancilla interactions that preserves the code distance~\cite{Dennis:2002,fowler2009high,PhysRevA.96.032341,PhysRevA.98.050301,chamberland2018flag,Yoder:2017b,Chamberland2020topo, PhysRevA.101.042312}. The time overhead can also be reduced by careful choice of sequential extractions ~\cite{delfosse2020beyond,delfosse2020short}.
Notably, as an alternative to cat-state measurements, flag error-correction gadgets~\cite{Chao2018,Chao2018few,Chamberland2018flagfaulttolerant,chao2020flag} circumvent the need of ancilla post-selection for arbitrary stabilizer codes,
while having a low qubit overhead for low-distance codes or large distance codes with low-weight stabilizer measurements. 

The extraction gadgets for Shor's scheme is arguably the smallest. As a trade-off, a large number of two-qubit gates are applied between data and ancilla qubits. 
For many quantum devices, two-qubit gates are usually the most challenging operations to be implemented with high fidelity~\cite{krantz2019quantum,bruzewicz2019trapped}. 
To minimize the data-ancilla interaction, Steane~\cite{steane1997active} and Knill~\cite{Knill2005nature} suggested to use transversal two-qubit gates to transfer the errors from the data block to an ancilla code block, then measure the ancilla block to gain error information.
Steane's scheme is specialized for CSS (Calderbank-Shor-Steane) codes~\cite{calderbank1996good,Steane:1996}, and needs two separate rounds of transversal two-qubit gates  for extracting the $X$- and $Z$-type stabilizers respectively.
Knill's scheme works for arbitrary stabilizer codes, and only needs one round of transversal gate to extract all the stabilizers. 
Using a constant number of Steane/Knill syndrome extraction, an arbitrary logical Clifford circuit can be implemented fault-tolerantly in $O(1)$ steps~\cite{Zheng_2020}.
Both Steane's and Knill's schemes are \textit{single-shot}, i.e., no repetition of measurements are required. Indeed, each data qubit is touched by $O(1)$ two-qubit gates. 
However, comparing to a cat state, the ancilla blocks 
for both Steane's and Knill's scheme are much harder to prepare, as they are as large as the data block~\cite{Steane:2003,paetznick2012,goto2016minimizing, PhysRevA.95.032339, PhysRevA.97.032331}.

A natural question to raise is 
whether it is possible to balance the 
complexities of ancilla preparation and data-ancilla interaction. 
In this work, we give a positive answer for CSS codes. 
We generate a family of extraction circuits including Shor's and Steane's construction as its two extremes. 
This family increases the complexity of ancilla construction in exchange for reducing the 
number of two-qubit gates between data and ancilla qubits required to fault-tolerantly measure the error.
Applying our construction on the toric code, we are able to use a single ancilla block to measure the plaquette operators ($Z$-stabilizer elements) inside any connected sublattice. 
In particular, one can partition the $L\times L$ toric lattice into patches, each of which contains $m\times m$ plaquettes.
Shor's and Steane's schemes correspond to the special cases $m=1$ and $m=L$, respectively. 
Moreover, by offsetting the partition periodically,  one can achieve fault tolerance within $O(L/m)$ measurement rounds. 
Indeed, the parameter $m$ simultaneously controls the ancilla block size and the time overhead of fault tolerance.
As a remark, our result is compatible with the fact that Shor's and Steane's schemes require $O(L)$ and $O(1)$ measurement rounds on the toric code respectively. 

Our paper is organized as follows. In Section~\ref{Sec:2}, we review the essential background and present our notation. In Section~\ref{Sec:3}, we discuss the construction of our family of extraction circuits. In particular, a subfamily called transversal gadgets are proposed. In Section~\ref{Sec:4} we discuss the application of transversal gadgets on the toric code, and analyze the time overhead of fault-tolerant syndrome measurements.
We conclude with some discussions in section~\ref{Sec:5}.

\section{Background}\label{Sec:2}

\subsection{Linear Codes}

We start by developing our notation for connecting sets and binary vector spaces. Given a set $\Omega$, the binary vector space spanned by $\Omega$ is defined as
\begin{equation}
    \ztwo[\Omega] := \left\{\sum_{v \in \phi} v \middle| \phi \subseteq \Omega, |\phi| < \infty\right\}.
\end{equation}
From the definition, we can see that there is a one-to-one correspondence between vectors of $\ztwo[\Omega]$ and finite subsets of $\Omega$. To simplify the notations, we do not distinguish between a finite subset $\phi \subseteq \Omega$ and the vector $\sum_{v\in\phi} v \in \ztwo[\Omega]$, i.e., we are allowed to write $\phi = \sum_{v\in\phi} v \in \ztwo[\Omega]$. This allows us to define unions and intersections of vectors, which will be helpful later.  
In particular, for each element $a \in \Omega$, $a$ has another two meanings: a vector $a \in \ztwo[\Omega]$ and a subset $\{a\} \subseteq \Omega$. 
The addition of two vectors in $\ztwo[\Omega]$ corresponds to the \textit{symmetric difference} of two finite subsets of $\Omega$, i.e., for any two vectors $A, B \subseteq \ztwo[\Omega]$, we have $A + B = (A \cup B) \setminus (A \cap B)$. For each finite subset $\phi \subseteq \Omega$, its cardinality $|\phi|$ is also referred to as the \textit{weight} of the vector $\phi\in \ztwo[\Omega]$.

The \textit{(standard) inner product} of $\ztwo[\Omega]$ is an $\ztwo$-bilinear form $$\langle\cdot,\cdot\rangle_{\Omega}: \ztwo[\Omega] \times \ztwo[\Omega] \rightarrow \ztwo$$ such that for any $a,b \in \Omega$, $\langle a,b\rangle_\Omega = 1$ if and only if $a = b$. Two vectors $\phi, \psi \in \ztwo[\Omega]$ are said to be \textit{orthogonal} if and only if $\langle \phi, \psi\rangle_\Omega = 0$.  More generally, two vector subspaces $V,W \subseteq \ztwo[\Omega]$ are said to be orthogonal (denoted by $V \bot W$) if we have $\langle v,w\rangle_\Omega = 0$ for every $v \in V$, $w \in W$. 

Let $\Omega$, $\Theta$ be two sets and $M: \ztwo[\Omega] \rightarrow \ztwo[\Theta]$ be a linear map. The kernel and image of $M$ is denoted by $\ker M$ and $\im M$. When $\Omega$ and $\Theta$ are finite,  the transpose of $M$ is defined to be a linear map $M^\transpose: \ztwo[\Theta] \rightarrow \ztwo[\Omega]$ such that
$$\langle M^\transpose a, b \rangle_\Theta = \langle a, M b \rangle_\Omega$$ for every $a \in \Omega$, $b \in \Theta$. In other words, under the standard bases $\Omega$ and $\Theta$, the matrix $A^\transpose$ is the transpose of the matrix $A$.

We now briefly discuss linear codes. 
Let $\Omega$ be a set of $n$ classical bits. 
A \textit{linear code} on $\Omega$ is a subspace $\code \subseteq \ztwo[\Omega]$, or equivalently an injective linear map $G: \ztwo^k \hookrightarrow \ztwo[\Omega]$ with $\im G = \code$ and  $k = \dim \code$. 
The vectors of $\code$ are called \textit{codewords}, and $G$ is referred to as the \textit{generator matrix}.
The \text{dual code} of $\code$ is defined by
\begin{eqnarray}
\code^\bot := \left\{ v \in \ztwo[\Omega]: \langle v, c\rangle = 0,\ \forall c \in \code \right\}.
\end{eqnarray}
Note that  $\code^\bot = \ker G^\transpose$. This is because for each $v \in \code^\bot$, we have
\begin{equation}
    \langle G^\transpose v, x\rangle = \langle v, Gx\rangle = 0
\end{equation}
for every $x \in \ztwo^k$, which indicates that $G^\transpose v = 0$.
The generator matrix of $\code^\bot$, denoted by $H$, is called the \textit{parity-check matrix} of $\code$ since $\code = \ker H^\transpose$.

\subsection{CSS Stabilizer Codes}

Let $\Omega$ be a set of $n$ qubits. The \textit{state space} of the quantum system $\Omega$ is denoted by $\Hilbert_\Omega = \mathbb{C}\left[\ztwo[\Omega]\right] \cong \mathbb{C}^{2^n}$. The \textit{Pauli group} on $\Omega$ is denoted by $\Pauligrp_\Omega$. For each qubit $q \in \Omega$, the Pauli $X$ and $Z$ operators on $\Hilbert_q$ is denoted by $X_q$ and $Z_q$. For each subset $\psi \subseteq \Omega$, the \textit{$X$-type} and \textit{$Z$-type} operators supporting on $\psi$ are defined by
\begin{equation}
X[\psi]:=\bigotimes_{q \in \Omega}X_q^{\langle\psi,q\rangle_\Omega} \in \Pauligrp_\Omega
\end{equation}
and
\begin{equation}
Z[\psi]:=\bigotimes_{q \in \Omega}Z_q^{\langle\psi,q\rangle_\Omega} \in \Pauligrp_\Omega
\end{equation}
respectively.

A \textit{quantum code} on $\Omega$ is a subspace $\code_Q \subseteq \Hilbert_\Omega$.
%A stabilizer of $\code$ is an operator $S$ on $\Hilbert_\Omega$ such that $S\ket{\psi} = \ket{\psi}$ for any codeword $\ket{\psi} \in \code$.
A \textit{logical operator} of $\code_Q$ is an operator $L$ on $\Hilbert_\Omega$ such that $L(\code_Q) \subseteq \code_Q$.
A \textit{stabilizer code} on $\Omega$ is defined by an abelian subgroup $\stabgrp \subseteq \Pauligrp_\Omega$ such that every operator $P \in \stabgrp$ has eigenvalues $\pm 1$, and $-\mathbbm{1}_{\Hilbert_\Omega} \notin \stabgrp$. The corresponding code space $\code_Q$ is the common $+1$ eigenspace of all the operators in $\stabgrp$. In particular, if $\dim \code_Q = 1$, the unique state in $\code_Q$ (up to a constant) is said to be a \textit{stabilizer state}. If $\stabgrp = \stabgrp_X \oplus \stabgrp_Z$, where elements in $\stabgrp_X$ ($\stabgrp_Z$) are all $X$-type ($Z$-type), we say that $\stabgrp = \stabgrp_X \oplus \stabgrp_Z$ is a Calderbank-Shor-Steane (CSS) code~\cite{calderbank1996good,Steane:1996}, and $\stabgrp_X$ and $\stabgrp_Z$ are the $X$- and $Z$-stabilizer, respectively.
We can represent $\stabgrp_X$ and $\stabgrp_Z$ by two binary vector subspaces
\begin{equation}
    \code_X := \left\{\psi \in \ztwo[\Omega]: X[\psi] \in \stabgrp_X\right\} \cong \stabgrp_X
\end{equation}
and
\begin{equation}
    \code_Z := \left\{\psi \in \ztwo[\Omega]: Z[\psi] \in \stabgrp_Z\right\} \cong \stabgrp_Z
\end{equation}
respectively.
Note that for any $\psi, \phi \subseteq \Omega$, $X[\psi]$ and $Z[\phi]$ commute with each other if and only if $\psi$ and $\phi$ are orthogonal. As $\stabgrp = \stabgrp_X \oplus \stabgrp_Z$ is abelian, we must have $\code_X \bot \code_Z$. In this paper, we prefer the binary vector space representation of CSS codes, and use the notation $\code_X \bot \code_Z$ to denote a CSS code.

For a CSS code $\code_X \bot \code_Z$ with a codespace $\code_Q \subseteq \Hilbert_\Omega$, the dimension of $\code_Q$ is $2^k$, where 
\begin{equation}
    k = n - \dim \code_X - \dim \code_Z.
\end{equation}
Thus we say that $\code_X \bot \code_Z$ encodes $k$ logical qubits.
For any vector $\psi \in \code_Z^\bot$, $X[\psi]$ commutes with any stabilizers and hence must be a logical operator. Similarly, for any $\phi \in \code_X^\bot$, $Z[\phi]$ is a logical operator. 
In fact, we are able to choose $k$ vectors $\psi_1, \cdots, \psi_k \in \code_Z^\bot$ and another $k$ vectors $\phi_1, \cdots, \phi_k \in \code_X^\bot$ such that
$\langle \psi_i, \phi_j \rangle = \delta_{ij}$. Defining $\bar{X}_i := X[\psi_i]$ and $\bar{Z}_j := Z[\phi_j]$, we have
\begin{equation}
\bar{X}_i \bar{Z}_j =  (-1)^{\delta_{ij}} \bar{Z}_j \bar{X}_i.
\end{equation}
Indeed, $\bar{X}_i$ and $\bar{Z}_i$ can be regarded as the logical Pauli $X$ and $Z$ operator of the $i$-th logical qubit, respectively.
%Indeed, we can regard $\bar{X}_i|_\code$ ($\bar{Z}_i|_\code$), the restriction of $\bar{X}_i$ ($\bar{Z}_i$) on $\code$, as the logical Pauli $X$ ($Z$) operator of the $i$-th logical qubit. 
%Note that $\bar{X}_i|_\code$ and $\bar{Z}_j|_\code$ generate $\textrm{End}\ \code$, the algebra of linear operators on $\code$, whose centralizer is generated by the identity $\mathbbm{1}_\code$. To prove that a unitary logical operator $L$ is the identity (up to a phase), it suffices to prove that $L|_\code$ commutes with every $\bar{X}_i|_\code$ and $\bar{Z}_j|_\code$. 

\section{Syndrome Extractions for CSS codes} \label{Sec:3}

On a stabilizer code, the errors are detected by measuring the stabilizer elements.
For CSS codes, it is natural to measure the $Z$- and $X$-stabilizer elements so that the $X$-errors (bit flips) and the $Z$-errors (phase flips) can be handled separately.
Here, we are regarding a $Y$-error as the combination of an $X$-error and a $Z$-error. 
In this paper, we focus on $Z$-stabilizer measurements, since an analysis of $X$-stabilizer measurements would be the same up to a Hadamard transform.

Let $\code_X \bot \code_Z$ be a CSS code on $\data$ with a codespace $\code_Q \subseteq \Hilbert_\Omega$. Suppose we have an $X$-error $X[\psi]$ ($\psi \subseteq \Omega$) and we are measuring the $Z$-stabilizer elements $\left\{Z[\phi_a]\right\}_{a \in \Lambda}$, where $\phi_a \in \code_Z$ and $\Lambda$ is a finite set of \textit{syndrome bits}. For each syndrome bit $a \in \Lambda$, the value of $Z[\phi_a]$, which is $(-1)^{\langle \psi, \phi_a \rangle_\Omega}$, uniquely determines the value $\langle \psi, \phi_a \rangle_\Omega \in \ztwo$. We assign the value of $a$ to be $\sigma_a = \langle \psi, \phi_a \rangle_\Omega$, and the vector 
\begin{equation}
\sigma := \sum_{a \in \Lambda} \sigma_a a \in \ztwo[\Lambda]
\end{equation} is called the \textit{syndrome}. 
The measurements can be represented by a linear map $H: a \mapsto \phi_a$. We have $\sigma = H^\transpose \psi$ from the following observation
\begin{equation}
    \sigma_a = \langle \sigma, a\rangle = \langle \psi, H a \rangle = \langle H^\transpose \psi, a \rangle.
\end{equation}
Informally speaking, $H$ is a parity check matrix of the linear code $\code_Z^\bot$. For our purposes, however, we do not require that $\im H = \code_Z$. In addition, redundant parity checks are allowed, i.e., $H$ is not necessarily injective. The matrix $H$ is referred to as a \textit{$Z$-check matrix} of $\code_X \bot \code_Z$, which is formally defined as follows:
\begin{Definition}[$Z$-check matrices]
A $Z$-check matrix of a CSS code $\code_X \bot \code_Z$ on $\data$ is a matrix $H: \ztwo[\Lambda] \rightarrow \ztwo[\data]$ such that $\im H \subseteq \code_Z$.
\end{Definition}

If we are at the end of the computation, the syndrome of a $Z$-check matrix $H$ can be obtained by measuring all qubits of $\Omega$ in the $Z$-basis: if $\Omega$ is measured to be in the state $\ket{m} \in \Hilbert_\Omega$, where $m \in \ztwo[\Omega]$, then the syndrome $\sigma = H^\transpose m$. In the intermediate steps, however, we
are not allowed to measure the data qubits directly, as the single-qubit $Z$-measurements anticommute with the $X$-stabilizer elements. 
A general idea shared by both Shor's and Steane's syndrome extraction protocols is to transfer the $X$-errors to a set of \textit{ancilla qubits} $\Theta$ by $\cnot$ gates: as $\cnot$ propagates the $X$ errors on the control qubit to the target qubit, we can perform $\cnot$ gates with controls in $\Omega$ and targets in $\Theta$, then apply $Z$-measurements on all ancilla qubits to detect these errors.
Of course, this is far from being a valid construction. In the following, we explore the necessary and sufficient conditions for a valid extraction circuit. To simplify our problem, we do not initially consider the challenge of fault tolerance.

 As a first step, we encode the information of the $\cnot$ gates by a matrix $\Gamma: \ztwo[\Theta] \rightarrow \ztwo[\Omega]$, where for each ancilla qubit $a \in \Theta$ and data qubit $d \in \Omega$, $\langle \Gamma a, d \rangle = 1$ if and only if a $\cnot$ gate with control $d$ and target $a$ is applied. As these $\cnot$ gates commute with each other and we do not consider the fault tolerance properties, the order of these $\cnot$ gates does not matter. The product of all $\cnot$ gates is denoted by $\network_\Gamma$.
 One can easily verify the following identities:
 \begin{eqnarray}
 \network_\Gamma X[\psi]  X[\psi'] &=& X[\psi] X[\psi' + \Gamma^\transpose \psi] \network_\Gamma\label{Xrelation}\\
 \network_\Gamma Z[\phi]  Z[\phi'] &=& Z[\phi+\Gamma \phi'] Z[\phi'] \network_\Gamma\label{Zrelation}
 \end{eqnarray}
 where $\psi,\phi \in \ztwo[\Omega]$ and $\psi',\phi' \in \ztwo[\Theta]$.
 
 Suppose the ancilla block $\Theta$ is prepared in a state $\ket{\theta} \in \Hilbert_\Theta$. When there are no errors on $\Omega$, the action of $\network_\Gamma$ is required to be trivial, i.e., we should have
\begin{equation}\label{logicalid}
    \network_\Gamma \ket{\omega}\ket{\theta} = \ket{\omega}\ket{\theta},\quad \forall \ket{\omega} \in \code_Q.
\end{equation}
Note that the code space $\code_Q$ is spanned by
\begin{equation}
    \left\{Z[\phi] \ket{\overline{+^k}}: \phi \in \code_X^\bot\right\},
\end{equation}
where $\ket{\overline{+^k}}$, the logical $\ket{+^k}$ state, is the CSS stabilzier state of $\code_Z^\bot \bot \code_Z$. This is true because we have enumerated all the logical $Z$-type operators. From (\ref{Zrelation}), $Z[\phi]$ commutes with $\network_\Gamma$ for any $\phi \in \ztwo[\Omega]$. Thus (\ref{logicalid}) can be simplified as
\begin{equation}
    \network_\Gamma \ket{\overline{+^k}}\ket{\theta} = \ket{\overline{+^k}}\ket{\theta}.\label{logicalplus}
\end{equation}

%Here, we exclude the possibilities that $\ket{\theta}$ evolves to another state $\ket{\theta'} \in \Hilbert_\Theta$ after applying $\network_\Gamma$.
%From (\ref{logicalid}) we can deduce some conditions for $\ket{\theta}$.
Let $X[\psi]$ ($\psi \subseteq \Omega$) be an $X$-error on a codeword $\ket{\omega} \in \code_Q$. From (\ref{Xrelation}), we have
\begin{eqnarray}
    &\ &\network_\Gamma \left(X[\psi]\ket{\omega}\right) \ket{\theta}= (X[\psi]\ket{\omega}) (X[\Gamma^\transpose\psi]\ket{\theta}).\label{propagation}
\end{eqnarray}
If $\ket{\theta}$ has a non-trivial $Z$-stabilizer represented by $\tilde{\code}_Z \subseteq \ztwo[\Theta]$,
fixing a $Z$-check matrix $\tilde{H}: \ztwo[\Lambda] \rightarrow \ztwo[\Theta]$ with $\im \tilde{H} \subseteq \tilde{\code}_Z$, the syndrome of $\tilde{H}$ will be $\tilde{H}^\transpose \Gamma^\transpose \psi \in \ztwo[\Lambda]$. If $H = \Gamma \tilde{H}$, then we can obtain the syndrome of $H$ by measuring all qubits of $\Theta$ in the $Z$-basis. One could naturally ask the following question:

\begin{Question}\label{Q1}
Given two matrices $\tilde{H}: \ztwo[\Lambda] \rightarrow \ztwo[\Theta]$ and $\Gamma: \ztwo[\Theta] \rightarrow \ztwo[\Omega]$ such that $H = \Gamma \tilde{H}$ is a $Z$-check matrix of $\code_X \bot \code_Z$, can we find a state $\ket{\theta} \in \Hilbert_\Theta$ such that $\network_\Gamma \ket{\overline{+^k}}\ket{\theta} = \ket{\overline{+^k}}\ket{\theta}$ and $Z[\phi] \ket{\theta} = \ket{\theta}$ for every $\phi \in \im 
\tilde{H}$? Or equivalently can the circuit 
\begin{center}
\begin{quantikz}
 \lstick{$\Omega$}&\qw\qwbundle{} &\gate[2]{\network_\Gamma}& \qw\qwbundle{}  &\\
& \lstick{$\ket{\theta}$} & \qwbundle{} &\meter{}\qwbundle{}
\end{quantikz}
\end{center}
extract the syndrome of $H$?
\end{Question}

\begin{figure*}
\begin{subfigure}{\linewidth}
\caption{}
\subfile{circuit2}
\end{subfigure}
\begin{subfigure}{\linewidth}
\caption{}

\begin{center}
\resizebox{\linewidth}{!}{
\begin{tabular}{c|c|c|c|c|c}
& Bare Ancilla & Cat State & Steane & Scheme A & Scheme B\\
\hline
$\tilde{H}^\transpose$ & $\left[1\right]$ & $\left[\begin{array}{cccc} 1 & 1&1&1\end{array}\right]$ & $\left[\begin{array}{ccccccc}
1 & 1 & 1 & 1 & 0 & 0 & 0\\
0 & 1 & 1 & 0 & 1 & 1 & 0\\
0 & 0 & 1 & 1 & 0 & 1 & 1
\end{array}\right]$ &
$\left[\begin{array}{cccccc}
1 & 1 & 1 & 1 & 0 & 0\\
0 & 1 & 1 & 0 & 1 & 1
\end{array} \right]$
& $\left[\begin{array}{ccc}
1 & 1 & 0\\
0 & 1 & 1
\end{array}\right]$\\
\hline
$\Gamma^\transpose$ & $\left[\begin{array}{ccccccc}
1&1&1&1&0&0&0\end{array}\right]$ & 
$\left[\begin{array}{c|c}
\mathbbm{I}_4 & 0_{4\times 3}
\end{array}\right]$ & 
$\mathbb{I}_7$
% $\left[\begin{array}{ccccccc}
% 1 & 0 & 0 & 0 & 0 & 0 & 0\\
% 0 & 1 & 0 & 0 & 0 & 0 & 0\\
% 0 & 0 & 1 & 0 & 0 & 0 & 0\\
% 0 & 0 & 0 & 1 & 0 & 0 & 0\\
% 0 & 0 & 0 & 0 & 1 & 0 & 0\\
% 0 & 0 & 0 & 0 & 0 & 1 & 0\\
% 0 & 0 & 0 & 0 & 0 & 0 & 1
% \end{array}\right]$ 
& 
$\left[\begin{array}{c|c}
\mathbbm{I}_6 & 0_{6\times 1}
\end{array}\right]$&  
$\left[\begin{array}{ccccccc}
1 & 0 & 0 & 1 & 0 & 0 & 0\\
0 & 1 & 1 & 0 & 0 & 0 & 0\\
0 & 0 & 0 & 0 & 1 & 1 & 0
\end{array}\right]$
\\
\hline
$H^\transpose$ & $\left[\begin{array}{ccccccc}
1&1&1&1&0&0&0\end{array}\right]$ & $\left[\begin{array}{ccccccc}
1&1&1&1&0&0&0\end{array}\right]$ & $\left[\begin{array}{ccccccc}
1 & 1 & 1 & 1 & 0 & 0 & 0\\
0 & 1 & 1 & 0 & 1 & 1 & 0\\
0 & 0 & 1 & 1 & 0 & 1 & 1
\end{array}\right]$ & 
$\left[\begin{array}{ccccccc}
1 & 1 & 1 & 1 & 0 & 0 & 0\\
0 & 1 & 1 & 0 & 1 & 1 & 0
\end{array}\right]$&
$\left[\begin{array}{ccccccc}
1 & 1 & 1 & 1 & 0 & 0 & 0\\
0 & 1 & 1 & 0 & 1 & 1 & 0
\end{array}\right]$\\
\hline
\begin{tabular}{c}
$Z$-stabilizer\\
generators\end{tabular} & $Z_{1'}$ & $Z_{1'}Z_{2'}Z_{3'}Z_{4'}$ & 
\begin{tabular}{c}
$Z_{1'}Z_{2'}Z_{3'}Z_{4'}$, $Z_{2'}Z_{3'}Z_{5'}Z_{6'}$,\\ $Z_{3'}Z_{4'}Z_{6'}Z_{7'}$
\end{tabular} &
$Z_{1'}Z_{2'}Z_{3'}Z_{4'}$, $Z_{2'}Z_{3'}Z_{5'}Z_{6'}$
& $Z_{1'}Z_{2'}$, $Z_{2'}Z_{3'}$\\
\hline
\begin{tabular}{c}$X$-stabilizer\\
generators
\end{tabular}& None & 
$X_{1'}X_{2'}$, $X_{2'}X_{3'}$, $X_{3'}X_{4'}$ 
& \begin{tabular}{c}
$X_{1'}X_{2'}X_{3'}X_{4'}$,
$X_{2'}X_{3'}X_{5'}X_{6'}$\\
$X_{3'}X_{4'}X_{6'}X_{7'}$, 
$X_{1'}X_{2'}X_{5'}$
\end{tabular} & 
\begin{tabular}{c}
$X_{1'}X_{2'}X_{3'}X_{4'}$,
$X_{2'}X_{3'}X_{5'}X_{6'}$\\
$X_{3'}X_{4'}X_{6'}$,
$X_{1'}X_{2'}X_{5'}$
\end{tabular}
&$X_{1'}X_{2'}X_{3'}$ 
\end{tabular}
}
\end{center}

\end{subfigure}
\caption{
For the Steane [[7,1,3]] code, we illustrate by (a) quantum circuits and (b) matrices how our division of $H$ into $\Gamma$ and $\tilde{H}$ enables us to describe three common ancilla blocks: bare qubits, cat states, and Steane ancilla. It also enables protocols that extracts two $Z$-stabilizer elements in parallel, as shown in Scheme A and Scheme B.
The $Z$- and $X$-stabilizer generators are generated from $\im \tilde{H}$ and $\ker \tilde{H}^\transpose$, respectively. 
The ancilla state for scheme B is a cat state $\ket{000}+\ket{111}$, which can be directly prepared without verification. However, 
as the circuit is not transversal on the ancilla block, we will require a flag decoding circuit~\cite{chao2020flag} to achieve fault tolerance.
}\label{fig1}
\end{figure*}

Suppose we already have a satisfying ancilla state $\ket{\theta}$ for a given $\Gamma$ and $\tilde{H}$.
In (\ref{propagation}), if we take $\psi \in \code_Z^\bot$ and combine (\ref{logicalplus}), we will have
\begin{eqnarray*}
\ket{\overline{+^k}}\otimes  \left(X[\Gamma^\transpose\psi]\ket{\theta}\right) = \ket{\overline{+^k}}\otimes \ket{\theta},
\end{eqnarray*}
i.e., $X[\Gamma^\transpose\psi]\ket{\theta} = \ket{\theta}$. 
Using $\tilde{\code}_X \subseteq \ztwo[\Theta]$ to represent the $X$-stabilizer of $\ket{\theta}$, $X[\Gamma^\transpose \psi] \ket{\theta} = \ket{\theta}$ is equivalent as $\Gamma^\transpose \psi \in \tilde{\code}_X$. Therefore, we must have
\begin{equation}
    \Gamma^\transpose\left(\code_Z^\bot\right) \subseteq \tilde{\code}_X.\label{czbotcx}
\end{equation}
On the other hand, let $\tilde{\code}_Z \subseteq \ztwo[\Theta]$ represents the $Z$-stabilizer group of $\ket{\theta}$. For every $\phi \in \tilde{\code}_Z$, we have
\begin{eqnarray}
& & \ket{\overline{+^k}} \ket{\theta} = \network_\Gamma \ket{\overline{+^k}} \left(Z[\phi] \ket{\theta}\right)= \left(Z[\Gamma\phi]\ket{\overline{+^k}}\right) \ket{\theta}\nonumber
\end{eqnarray}
Therefore $\Gamma\phi \in \code_Z$ and hence
\begin{equation}
    \Gamma\left(\tilde{\code}_Z\right) \subseteq \code_Z. \label{czcz}
\end{equation}
The conditions (\ref{czbotcx}) and (\ref{czcz}) implies that $\network_\Gamma$ preserves the CSS stabilizer group
\begin{eqnarray}
\left(\code_Z^\bot \oplus \tilde{\code}_X\right) \bot \left(\code_Z \oplus \tilde{\code}_Z\right).\label{stabilizer}
\end{eqnarray}
If $\tilde{\code}_X = \tilde{\code}_Z^\bot$, i.e., $\ket{\theta}$ is a CSS stabilizer state and hence $\ket{\overline{+^k}}\ket{\theta}$ is the stabilizer state of (\ref{stabilizer}), the condition (\ref{logicalplus}) must hold.

Our problem now becomes finding a CSS stabilizer state $\ket{\theta}$ of $\tilde{\code}_X \bot \tilde{\code}_Z$ ($\tilde{\code}_X = \tilde{\code}_Z^\bot$) satisfying (\ref{czbotcx}) and (\ref{czcz}). Of course, we must have $\im \tilde{H} \subseteq \tilde{\code}_Z$ and hence $\tilde{\code}_X \subseteq \left(\im \tilde{H}\right)^\bot$. 
Therefore, $\ket{\theta}$ exists if and only if $\tilde{\code}_Z = \im \tilde{H}$, $\tilde{\code}_X = \tilde{\code}_Z^\bot$ satisfy (\ref{czbotcx}) and (\ref{czcz}). In fact, this is always true by the following lemma:
\begin{Lemma}
$\Gamma(\im \tilde{H}) \subseteq \code_Z$ and $\Gamma^\transpose\left(\code_Z^\bot\right) \subseteq \left(\im \tilde{H}\right)^\bot$.
\end{Lemma}
\begin{proof}
The first part follows from a direct calculation
\begin{equation*}
    \Gamma(\im \tilde{H}) = \Gamma \tilde{H}(\ztwo[\Lambda]) = H(\ztwo[\Lambda]) \subseteq \code_Z.
\end{equation*}
To prove the latter part, for any $\psi \in \code_Z^\bot$ and $a \in \Lambda$, we have
\begin{eqnarray}
\langle \Gamma^\transpose \psi, \tilde{H}a \rangle_\Theta = \langle \psi, \Gamma\tilde{H}a  \rangle_\Omega = \langle \psi, H a  \rangle_\Omega = 0
\end{eqnarray}
as $Ha \in \code_Z$. Thus $\Gamma^\transpose \psi \in \left(\im \tilde{H}\right)^\bot$.
\end{proof}

From the discussions above, we can conclude that given a $Z$-check matrix $H$ of $\code_X \bot \code_Z$, any decomposition $H = \Gamma \tilde{H}$ corresponds a valid gadget extracting the syndrome of $H$. All the information of the gadget is determined by $\Gamma$ and $\tilde{H}$, which allows us to define a gadget in an abstract way:
\begin{Definition}[$Z$-extraction gadgets]
A $Z$-extraction gadget of a CSS code $\code_X \bot \code_Z$ on $\Omega$ is a tuple $(\Theta, \Lambda, \Gamma, \tilde{H})$, where $\Theta$ is the set of ancilla qubits, $\Lambda$ is the set of syndrome bits, $\tilde{H}: \ztwo[\Lambda] \rightarrow \ztwo[\Theta]$ and $\Gamma: \ztwo[\Theta] \rightarrow \ztwo[\Omega]$ are two matrices such that $H = \Gamma \tilde{H}$ is a $Z$-check matrix of $\code_X \bot \code_Z$.
$\Gamma$, $\tilde{H}$ and $H$ are referred to as the \textbf{gate matrix}, \textbf{ancilla check matrix} and \textbf{data check matrix}, respectively.
The ancilla block $\Theta$ is prepared in the CSS state of $\left(\im \tilde{H}\right)^\bot \bot \im \tilde{H}$, while the gate applied between $\Omega$ and $\Theta$ is $\network_\Gamma$.
\end{Definition}

If we apply two gadgets 
$\gadget_1 = (\Theta_1, \Lambda_1, \Gamma_1, \tilde{H}_1)$
and $\gadget_2 = (\Theta_2, \Lambda_2, \Gamma_2, \tilde{H}_2)$, we can view them as an united gadget $\gadget = (\Theta, \Lambda, \Gamma, \tilde{H})$ such that $\Theta = \Theta_1 \cup \Theta_2$, $\Lambda = \Lambda_1 \cup \Lambda_2$. The gate matrix $\Gamma$ is defined by 
\begin{equation}
    \Gamma a = \left\{\begin{array}{lr}
    \Gamma_1 a & \quad a \in \Theta_1,\\
    \Gamma_2 a & \quad a \in \Theta_2.
    \end{array} \right.
\end{equation}
and the ancilla check matrix $\tilde{H}$ is defined by
\begin{equation}
    \tilde{H} c = \left\{\begin{array}{lr}
    \tilde{H}_1 c & \quad c \in \Lambda_1,\\
    \tilde{H}_2 c & \quad c \in \Lambda_2.
    \end{array} \right.
\end{equation}
We say that $\gadget$ is a sum of $\gadget_1$ and $\gadget_2$, denoted by $\gadget = \gadget_1 \oplus \gadget_2$.

We now review Shor's and Steane's constructions in our construction:
\begin{Example}[Shor's scheme~\cite{shor1996fault,divincenzo1996fault}]
In Shor's scheme, 
each syndrome bit $c \in \Lambda$ corresponding to the $Z$-stabilizer element $Z[\phi_c]$ on the data block $\Omega$ is extracted by a separate gadget $\gadget_c = \left(\Theta_c, \{c\}, \Gamma_c, \tilde{H}_c\right)$ such that $\Gamma_c\tilde{H}_c (c) = \phi_c$.
The whole gadget is therefore
$\gadget = \bigoplus_{c \in \Lambda} \gadget_c$, and we say that $\gadget$ is a Shor-style gadget.

The simplest choice of $\gadget_c$ is to set $\Theta_c = \{b\}$ to have one ancilla qubit, and $\tilde{H}_c(c) = b$, $\Gamma_c(b) = \phi_c$.
Since $\im \tilde{H_c} \cong \ztwo$ and $\left(\im \tilde{H_c}\right)^\bot = 0$, the  ancilla qubit $b$ is stabilized by the single-qubit Pauli $Z$ operator. 
This is known as the \textbf{bare-ancilla gadget}. If all the $\gadget_c$s are bare-ancilla, the gate matrix of $\gadget$ is identical to $H$, while the ancilla check matrix is identical to the identity matrix of dimension $|\Lambda|$.

Another choice of $\gadget_c$ is to set $|\Theta_c| = |\phi_c|$. 
We fix a set bijection $\gamma: \Theta_c \leftrightarrow \phi_c$, and define
the gate matrix $\Gamma_c$ by $\Gamma_c(b) = \gamma(b)$ for every $b \in \Theta_c$. The 
ancilla check matrix is defined by $\tilde{H}_c(c) = \Theta_c \in \ztwo[\Theta_c]$.
One can verify that the ancilla state is the cat state $\ket{+^{|\phi_c|}} + \ket{-^{|\phi_c|}}$. This is known as the \textbf{cat-state gadget}. 

\end{Example}

\begin{Example}[Steane's scheme~\cite{steane1997active}]
In Steane's scheme, all the $Z$-stabilizer elements are extracted by one ancilla block. The data check matrix $H: \ztwo[\Lambda] \rightarrow \ztwo[\Omega]$ is the generator matrix of $\code_Z$, or the parity check matrix of $\code_Z^\bot$. 
The ancilla block $\Theta$ is identical to the data block $\Omega$. The gate matrix $\Gamma$ is identity matrix under the standard bases, 
and the ancilla check matrix $\tilde{H}$ is identical to $H$. The ancilla state is the CSS stabilizer state of $\tilde{\code}_Z^\bot \bot \tilde{\code}_Z$, which is the logical $\ket{+^k}$ state, where $k$ is number of logical qubits.
\end{Example}

We take Steane's $[[7,1,3]]$ code as an example to illustrate these standard methods in FIG.~\ref{fig1}. In addition, we illustrate a non-fault-tolerant circuit that uses $3$ ancilla qubits to extract two weight-$4$ $Z$-stabilizer elements in parallel, referred to as Scheme A and Scheme B. 

In fault-tolerant error correction, when we apply the extraction gadgets onto the data block, we need to consider the errors introduced by the gadgets themselves.
Clearly, not every gadget $(\Theta,\Lambda,\Gamma,\tilde{H})$ can be directly used for fault-tolerant error correction. For example, if there are multiple $\cnot$ gates on an ancilla qubit, correlated errors could be generated on the data block. Techniques from flag error-correction~\cite{Chao2018,Chao2018few,chao2018quantum,chao2020flag} could be applied to detect the crucial errors on the ancilla block. To keep our discussion simple, however, we focus our attention to a family of gadgets having the following ``transversal'' property inspired from Shor's and Steane's schemes:
\begin{Definition}[Transversal Gadgets]
A $Z$-extraction gadget $(\Theta, \Lambda, \Gamma, \tilde{H})$ is said to be \textit{transversal}, if $|\Gamma(a)|$ = 1 for every $a \in \Theta$, i.e., $\network_\Gamma$ is transversal on the ancilla block $\Theta$. As a remark, the $\cnot$ gates of a transversal gadget could be non-transversal on the data block. 
\end{Definition}
Assuming the ancilla block can be prepared fault-tolerantly in the sense that no correlated errors can occur on the block, a transversal gadget will not introduce correlated errors to the data block. A way to construct transversal gadgets is to start from Steane's scheme and split each ancilla qubit as a set of qubits, which is formally described in Procedure 1.

\begin{algbox}[H]
\caption{Construction of a Transversal Gadget}\label{alg1}
\begin{algorithmic}[1]
\State Pick a $Z$-check matrix $H: \ztwo[\Lambda] \rightarrow \ztwo[\Omega]$.
\State For each $a \in \Omega$, decompose $\psi_a = H^\transpose a \subseteq \Lambda$ as
\begin{equation*}
    \psi_a = \bigcup_{b \in \Theta_a} \phi_b,
\end{equation*}
for some set of labels $\Theta_a$ and disjoint non-empty subsets $\{\phi_b\}_{b\in \Theta_a}$ of $\psi_a$.  For two different $a,a'\in \Omega$, $\Theta_a \cap \Theta_{a'} = \emptyset$.
\State Let $\Theta = \bigcup_a \Theta_a$. 
\State Let $\Gamma: \ztwo[\Theta] \rightarrow \ztwo[\Omega]$ be the map such that for every $b \in \Theta_a \subseteq \Theta$, $\Gamma b = a$. Indeed, $\Gamma^\transpose a = \Theta_a$.
\State Let $\tilde{H}: \ztwo[\Lambda] \rightarrow \ztwo[\Theta]$ be the map such that for each $b \in \Theta$, $\tilde{H}^\transpose b = \phi_b$.
\State Return $(\Theta,\Lambda,\Gamma,\tilde{H})$.
\end{algorithmic}
\end{algbox}

In Procedure~\ref{alg1}, the number of ancilla qubits $|\Theta|$ is at least $|\Omega|$ and at most $\sum_{a \in \Omega} |H^\transpose a|$.
In the former case, we do not make any split at all so that our gadget is kept to be Steane-style.
In the latter case, however, we make as many split as possible so that for every ancilla qubit $b \in \Theta$, $|\tilde{H}^\transpose b| = 1$. Equivalently, 
for each two different syndrome bits $c,c' \in \Lambda$, $\tilde{H}c \cap \tilde{H}c' = \emptyset$. This is to say that the gadget is Shor-style, as it can be decomposed as a sum of cat-state gadgets. 

We have not yet considered the measurement errors on the gadgets. In addition, as the $\cnot$ gates are not necessarily transversal on the data block, data errors on the same data qubit could lead to different syndromes. 
Shor's scheme handles these errors by repeating syndrome extractions~\cite{shor1996fault}, while Steane's scheme, however, does not require repetitions~\cite{steane1997active}.
For other transversal gadgets, we naturally expect that their behavior is somewhere between Shor-style and Steane-style gadgets. 
In the next section, we will study the transversal gadgets for the toric code~\cite{kitaev2003fault} and make some quantitative arguments. 

\section{Transversal Gadgets on the Toric Code}\label{Sec:4}

We briefly review the construction of the toric code~\cite{kitaev2003fault}. A toric code is a CSS code defined on an $L\times L$ periodic lattice on the torus. 
The lattice has a set $V$ of $L^2$ vertices, a set $E$ of $2L^2$ edges and a set $F$ of $L^2$ faces. 
Define the \textit{boundary map} 
\begin{eqnarray}
    \partial:  \ztwo[F] & \mapsto & \ztwo[E],\nonumber\\
     F \ni f & \mapsto & \left\{e \in E \middle| e\textrm{ borders }f \right\}, %\quad \forall f \in F
\end{eqnarray}
and the \textit{coboundary map}
\begin{eqnarray}
    \delta:  \ztwo[V] & \mapsto & \ztwo[E],\nonumber\\
     V \ni v & \mapsto & \left\{e \in E \middle| e\textrm{ is incident to }v \right\}.
\end{eqnarray}
For each $f \in F$ and $v \in V$, since $|\partial f \cap \delta v| = 0$ or $2$, we must have $\langle \partial f, \delta v\rangle = 0$. Indeed, $\im \delta \bot \im \partial$. Taking $\code_X = \im \delta$ and $\code_Z = \im \partial$, $\code_X \bot \code_Z$ is a well-defined CSS code on the edge set $E$. A non-trivial logical $X$-type ($Z$-type) operator is represented by a non-contractible loop (cut) on the torus, which has minimum length $L$. In other words, the toric code has distance $L$.

%The logical $Z$-type operators are the equivalence classes of $1$-cycles up to $1$-boundaries. The logial $Z$-type subgroup is \textit{the first homology group} $H_1(\chain) := \ker \partial_1 / \im \partial_2$.
%The logical $X$-type subgroup is \textit{the first cohomology group} $H^1(\chain) := \ker \delta_2 / \im \delta_1$. The the number of logical qubits is given by $\dim H_1(\chain) = \dim H^1(\chain)$\footnote{This is guaranteed by the universal coefficient theorem~\cite{}, which states that $H^i(\chain)$ is canonically isomorphic to $\Hom\left(H_i(\chain), \ztwo\right)$.}. 
%{\color{red} $H_1(\chain) \cong H_1(T)$.}

%A conventional choice of $Z$-extraction gadget is the bare-ancilla extraction~\cite{fowler2012surface}, represented by the tuple $(F, F, \partial, \mathbbm{1}_{F})$.
We consider using Procedure~\ref{alg1} to construct a transversal gadget for the $Z$-check matrix $\partial$, denoted by 
$(\tilde{E}, F, \Gamma, \tilde{\partial})$. For each qubit (edge) $e \in E$, $\partial^\transpose e$ contains exactly two faces, denoted by $f_1, f_2 \in F$.
If we choose to split $e$, it must be split into two qubits $\tilde{e}_1$ and $\tilde{e}_2$ such that $\tilde{\partial}^\transpose \tilde{e}_i = f_i$ and  $\Gamma(\tilde{e}_i) = e$. Geometrically, this procedure can be understood as cutting the torus along some chosen edges and obtaining a new topological space. The new space will have an edge set $\tilde{E}$ %a vertex set $\tilde{V}$, 
and a boundary map $\tilde{\partial}: \ztwo[F] \rightarrow \ztwo[\tilde{E}]$. 
Note that each split edge $\tilde{e} \in \tilde{E}$ is on the boundary of the obtained topological space. 
As the torus does not have a boundary, these split edges can be characterized by the subset $\tilde{\partial} F \subseteq \tilde{E}$. 
Moreover, if we cut the torus into several connected components so that the face set $F$ is decomposed as $F = \bigcup_{i} F_i$, where each $F_i$ is the face set of a component, then the gadget 
$(\tilde{E}, F, \Gamma, \tilde{\partial})$ can be decomposed as
\begin{equation}
(\tilde{E}, F, \Gamma, \tilde{\partial}) = 
    \bigoplus_i \left(\tilde{\partial}F_i,
    F_i, \Gamma|_{\tilde{\partial}F_i}, \tilde{\partial}|_{F_i}\right).
\end{equation}
That is, the ancilla block $\tilde{E}$ is decomposed as sub-blocks $\tilde{\partial}F_i$, and two different blocks are not entangled.
As a remark, if the torus is not being cut at all, $(\tilde{E},F,\Gamma,\tilde{\partial})$ is the Steane-style gadget; if the torus is cut into $L^2$ disjoint faces, $(\tilde{E},F,\Gamma,\tilde{\partial})$ is the Shor-style gadget. 

\subsection{Errors in Spacetime}

We now describe our model of fault-tolerant error correction.
As $X$- and $Z$-errors can be corrected separately, we can ignore the $Z$-errors on the data qubits and the $X$-stabilizer measurements.
Suppose our computation starts from time $0$ and never ends. We extract the $Z$-check matrix $\partial$ at every positive integer time.
For every $t \in \mathbb{Z}_+$, each data qubit could have an $X$-error
at time $t-1/2$, and the measurement outcome of each ancilla qubit at time $t$ could have a classical bit-flip error.
For now, we ignore the data errors between two $\cnot$s in the same extraction round. 

Suppose the transversal gadget applied at time $t \in \mathbb{N}$ is $\gadget_t = (\tilde{E}_t, F, \Gamma_t, \tilde{\partial}_{t})$, where $\Gamma_t\tilde{\partial}_{t} = \partial$. 
An $X$-error on the data qubit $e \in E$ at time $t-1/2$ is denoted by the pair $(e,t)$. The measurement error on the ancilla qubit $\tilde{e} \in \tilde{E}_t$ at time $t \in T$ is denoted by the pair $(\tilde{e},t)$.
The set of all single data-qubit errors, denoted by $D$, 
is
\begin{eqnarray}
D = E \times \mathbb{N},
\end{eqnarray}
while the set of measurement errors, denoted by $M$,  is
\begin{equation}
M = \bigsqcup_{t \in \mathbb{N}} \tilde{E}_t := \bigcup_{t \in \mathbb{N}} \tilde{E}_t \times \{t\}.
\end{equation}
An \textit{error history} is defined to be a finite subset $\psi \subseteq D \cup M$, or equivalently a vector $\psi \in \ztwo[D \cup M]$. 
%Given an error history $\psi$, the subset of errors of $\psi$ in the time interval $(0,t]$ is denoted by $\psi_t \subseteq \psi$.
The evaluation of data errors at time $t-1/2$ ($t \in \mathbb{N}$) is a map
\begin{eqnarray}
\dataerr_t: \ztwo[D \cup M] &\mapsto& \ztwo[E],\nonumber\\
\psi &\mapsto& \left\{ e:  (e,t) \in \psi \cap D \right\}.
\end{eqnarray}
For later convenience, the time coordinates of the errors are discarded. 
Similarly, the evaluation of measurement errors at time $t$ is a map
\begin{eqnarray}
\measerr_t: \ztwo[D\cup M]&\mapsto& \ztwo[\tilde{E}_t],\nonumber\\
\psi &\mapsto& %\sum_{\left(\tilde{e},t\right) \in \psi \cap M} \tilde{e}
\left\{\tilde{e}: (\tilde{e}, t) \in \psi \cap M\right\}.
\end{eqnarray}
The data error propagating to the the ancilla block $\tilde{E}_t$ is the accumulation of all data errors before time $t$, which can be evaluated by the map
\begin{equation}
    \bar{\dataerr}_t := \sum_{t'\le t} \dataerr_{t'}.
\end{equation} 
In particular, we define $\bar{\dataerr} := \sum_{t \in \mathbb{N}} \dataerr_t$.
The syndrome at time $t$ can be therefore evaluated by the map
\begin{eqnarray}
\sigma_t = \tilde{\partial}_{t}^\transpose\left(\Gamma^\transpose_t\bar{\dataerr}_t + \measerr_t\right) = \partial^\transpose \bar{\dataerr}_t + \tilde{\partial}_{t}^\transpose \measerr_t.
\end{eqnarray}
We can take the difference of the syndrome sequence $\{\sigma_t\}_{t \in \mathbb{N}}$, 
\begin{eqnarray}
\Delta_t &:=& \sigma_t - \sigma_{t-1}\nonumber\\
&=& \partial^\transpose\bar{\dataerr}_t + \tilde{\partial}_{t}^\transpose \measerr_t - \partial^\transpose \bar{\dataerr}_{t-1} - \tilde{\partial}_{t-1}^\transpose \measerr_{t-1}\nonumber\\
&=& \partial^\transpose\left(\bar{\dataerr}_t - \bar{\dataerr}_{t-1}\right) 
+ \tilde{\partial}_{t}^\transpose \measerr_t
- \tilde{\partial}_{t-1}^\transpose \measerr_{t-1}\nonumber\\
&=& \partial^\transpose \dataerr_t + \tilde{\partial}_{t}^\transpose \measerr_t
- \tilde{\partial}_{t-1}^\transpose \measerr_{t-1}\nonumber\\
&=& \partial^\transpose \dataerr_t + \tilde{\partial}_{t}^\transpose \measerr_t
+ \tilde{\partial}_{t-1}^\transpose \measerr_{t-1}.\label{deltat}
\end{eqnarray}
In the above calculation, we set $\sigma_0 = 0$, $\bar{\dataerr}_0 = 0$ and $\measerr_0 = 0$ for convenience.
We can see that the data error $\dataerr_t$ only contributes to $\Delta_t$, while the measurement error $\measerr_t$ contributes to both $\Delta_t$ and $\Delta_{t+1}$. The \textit{syndrome history} is defined as a map
\begin{eqnarray}
    \Sigma: \ztwo[D \cup M] &\mapsto& \ztwo[F \times \mathbb{N}]\nonumber\\
    \psi &\mapsto& \bigsqcup_{t \in \mathbb{N}} \Delta_{t}\psi.
\end{eqnarray}

We now analyze the behavior of each single error. For each data error $(e,t)$, one can verify that 
\begin{equation}
    \Sigma (e,t) = \left(\partial^\transpose e\right) \times \{t\} = \{f_1, f_2\} \times \{t\},
\end{equation}
where $f_1, f_2 \in F$ are the two faces sharing $e$ as their borders. 
For each measurement error $(\tilde{e}, t) \in M$, where $\tilde{e} \in \tilde{E}_t$, one can verify that
\begin{equation}
    \Sigma (\tilde{e}, t) = \left(\tilde{\partial}_{t}^\transpose \tilde{e} \right) \times \{t,t+1\}.
\end{equation}
If $\tilde{e}\in \tilde{\partial}_t F$, i.e., $\tilde{e}$ is an split edge, $\tilde{\partial}_{t}^\transpose \tilde{e}$ has only one face, and we say that $(\tilde{e}, t)$ is a \textit{type-I} error. Otherwise, $|\tilde{\partial}_{t}^\transpose \tilde{e}| = 2$, and $(\tilde{e}, t)$ is said to be a \textit{type-II} error. The set of type-I errors, denoted by $M_1$, is
\begin{equation}
    M_1 = \bigsqcup_{t\in \mathbb{N}} \tilde{\partial}_t F,
\end{equation}
while the set of type-II errors is denoted by $M_2 = M - M_1$.

The syndrome bit flips can be viewed as defects in the $(2+1)$-dimensional lattice:
each data error creates two defects in the same time slice; each type-I error creates two defects on the same location, but in two consecutive time slices; each type-II error creates four defects.
If $\gadget_t$ is a Shor-style gadget, all measurement errors at time $t$ will be of type-I, and the data and measurement errors are referred to as \textit{space-like} and \textit{time-like} errors respectively~\cite{Dennis:2002}.
If $\gadget_t$ is a Steane-style gadget, however, all measurement errors at time $t$ will be of type-II. 

\subsection{Minimum-Weight Perfect Matching Decoder}

Given an error history $\psi \in \ztwo[D\cup M]$ with observed syndrome history $\Sigma \psi \in \ztwo[F \times \mathbb{N}]$, a decoder \begin{equation}
    \decoder: \ztwo[F \times \mathbb{N}] \rightarrow \ztwo[D \cup M]
\end{equation} will take $\Sigma \psi$ as the input and output an estimation of error history $\psi' = \decoder\left(\Sigma\psi\right)$ such that $\Sigma\psi'=\Sigma\psi$. The optimal choice of the decoder is the minimum-weight-error (MWE) decoder $\mwedecoder$ defined by
\begin{equation}
    \mwedecoder(\Sigma\psi) = \argmin\limits_{\Sigma\psi'=\Sigma\psi} |\psi'|.
\end{equation}
If the gadgets are all Shor-style so that $M = M_1$, 
the decoding problem can be visualized by a \textit{decoder graph} 
$G$ with vertex set $F\times \mathbb{N}$ and edge set $D \cup M_1$: 
each $(f,t) \in F \times \mathbb{N}$ is a vertex and each error $\psi \in D \cup M_1$ is an edge connecting the two vertices (defects) in $\Sigma\psi$. 
The error histories are also called \textit{error chains}, as they can be visualized as sums of chains in $G$. Note that the map $\Sigma$ evaluates the boundary of a given error chain.
Decoding a syndrome $\Sigma\psi$ is essentially finding an error chain $\psi'$ whose boundary coincides with $\Sigma\psi$. As an error chain $\psi$ always matches the defects in $\Sigma\psi$ into pairs, the MWE decoder returns a minimum-weight perfect matching (MWPM) of the defects~\cite{Dennis:2002,fowler2009high,fowler2012surface}.

The existence of type-II errors complicates our problem, as they create four defects instead of two. We have to use hyper-edges to represent these errors in the decoder (hyper-)graph. 
Although we can still use a MWE decoder, the geometric meaning will be less clear. 
Notice that for any type-II error $(\tilde{e},t) \in M_2$ is equivalent to two data errors $(\Gamma_t\tilde{e},t) + (\Gamma_t\tilde{e}, t+1) \in \ztwo[D]$. As an approximation, we can pretend that the type-II errors do not exist, and use the MWPM decoder
\begin{equation*}
    \mwpmdecoder: \ztwo[F\times \mathbb{N}] \rightarrow \ztwo[D \cup M_1]
\end{equation*}
on the decoder graph with an edge set $D \cup M_1$.
For convenience, we define a map
\begin{eqnarray}
    \Pi: \ztwo[D \cup M] & \rightarrow & \ztwo[D \cup M_1],\\
    D \cup M_1  \ni \psi&\mapsto & \psi,\nonumber\\
    M_2  \ni (\tilde{e},t) &\mapsto & (\Gamma_t \tilde{e}, t) + (\Gamma_t \tilde{e}, t+1),\nonumber
\end{eqnarray}
that projects all the errors onto the decoder graph.
For each error history $\psi \in \ztwo[D \cup M]$, 
the MWPM decoder will regard it as an error chain $\Pi \psi$ on the decoder graph with total length $|\Pi\psi| = |\psi| + |\psi \cap M_2|$, or more explicitly
$$|\psi\cap D| + |\psi\cap M_1| + 2|\psi \cap M_2|.$$
Indeed, the MWPM decoder can only guarantee that
\begin{eqnarray}
    \left|\mwpmdecoder(\Sigma\psi)\right| \le |\Pi\psi| = |\psi| + |\psi \cap M_2|.\label{mwpmbound}
\end{eqnarray}
However, as shown below, $\mwpmdecoder$ can preserve the code distance $L$:
\begin{Theorem}\label{thm:1}
If $|\psi| < L/2$, $\mwpmdecoder$ will correct $\psi$ without introducing a logical error.
\end{Theorem}
\begin{proof}
Let $\psi' = \mwpmdecoder\left(\Sigma\psi\right)$. By applying the correction $\psi'$, the data error will become an undetectable error $\bar{\dataerr}\left(\psi+\psi'\right)$. 
Since the toric code has distance $L$, it suffices to show that 
$\left|\bar{\dataerr}\left(\psi+\psi'\right)\right| < L.$

From (\ref{mwpmbound}) and the fact that $|\psi' \cap D| \le |\psi'|$, we obtain $|\psi' \cap D| \le |\psi| + |\psi \cap M_2|$. Moreover,
\begin{eqnarray}
    & & |\psi' \cap D| + |\psi \cap D|\nonumber\\
    &\le& |\psi| + |\psi \cap D| + |\psi \cap M_2|\nonumber\\
    &\le& 2 |\psi| < L.\label{bound1}
\end{eqnarray}
The theorem is proved by combining (\ref{bound1}) and the fact that
\begin{eqnarray}
\left|\bar{\dataerr}\left(\psi+\psi'\right)\right| \le |\psi' \cap D| + |\psi \cap D|.
\end{eqnarray}
\end{proof}

\subsection{Fault-tolerant Error Correction in Finite Time}
In practice, the circuit will always end in some finite time $T$. All the errors and defects can only occur before $T$. Indeed, the MWPM decoder will have a finite decoder graph of size $O(TL^2)$. As the MWPM algorithm runs in polynomial time to the input size, for the purpose of reliable quantum memory, we can delay the decoding until the end of the circuit execution. However, this will be impractical for quantum computation tasks with non-Clifford gates~\cite{terhal2015quantum}.
Instead, we need to process the syndrome bits and correct the errors as soon as possible. As the information provided by the latest syndrome bits is always unreliable, they will be processed only when further syndrome bits are gathered.
For example, we can divide the time axis by some chosen time $$1=t_1 < t_2 < \cdots < t_i < \cdots.$$ In the $i$-th round of error correction, we decode the syndrome bits in the time interval $[t_i,t_{i+2})$ but only correct the errors in $[t_i-1/2,t_{i+1}-1/2)$; then discard the syndrome bits in $[t_i,t_{i+1})$ while keeping the syndrome bits in $[t_{i+1},t_{i+2})$ for the next round of correction. Note that the syndrome bits at time $t_{i+1}$ need to be updated. This is known as the \textit{overlapping recovery method}~\cite{Dennis:2002,delfosse2020beyond}. The details are formally described in Procedure~\ref{recovery}, for which we define
\begin{equation}
    \psi[t,t'] = \left\{ (e,t'') \in \psi \middle| t\le t''< t'\right\}
\end{equation}
and
\begin{equation}
    \Sigma[t,t']\psi = \bigsqcup_{t \le t'' < t'} \Delta_{t''} \psi
\end{equation}
for convenience.
\begin{algbox}[H]
\caption{Overlapping Recovery}\label{recovery}
\begin{algorithmic}[1]
\State $i \leftarrow 1$.
\State Use MWPM to find an error history $\psi'$ such that $$\Sigma[t_i,t_{i+2}]\left(\Pi\psi+\psi'\right) = 0.$$
\State $\psi \leftarrow \psi + \psi'[t_i,t_{i+1}]$.
\State $i \leftarrow i+1$, goto 2.
\end{algorithmic}
\end{algbox}
In the $i$-th iteration of Procedure~\ref{recovery}, 
the decoder graph for MWPM only contains the vertices (syndrome bits) and edges (errors) in $[t_i-1/2,t_{i+2}-1/2)$. In particular, the type-I errors at time $t_{i+2}-1$ will become open edges, i.e., edges connecting to the time boundary. The defects not only can be paired with each other, but also can be fused with the time boundary so that its lifetime is extended to the next round. Intuitively, if the distance from time slice $t_i$ to $t_{i+1}$ on the decoder graph, denoted by $d(t_i,t_{i+1})$, is too small, it would be too easy for the decoder to fuse a defect at time $t_i$ to the time boundary. As a consequence, errors can hardly be corrected.
If we use the Shor-style gadget for all the time, we will have $d(t_i,t_{i+1})=|t_{i+1}-t_i|$, and it has been shown that $|t_{i+1}-t_i| = O(L)$ suffices for fault-tolerance~\cite{Dennis:2002}. Naturally, the result should be generalized as $d(t_i,t_{i+1}) = O(L)$ for arbitrary choices of gadgets. To show this, we prove the following:

\begin{figure*}[htbp]
\begin{subfigure}{0.48\textwidth}
\includegraphics[width=\linewidth]{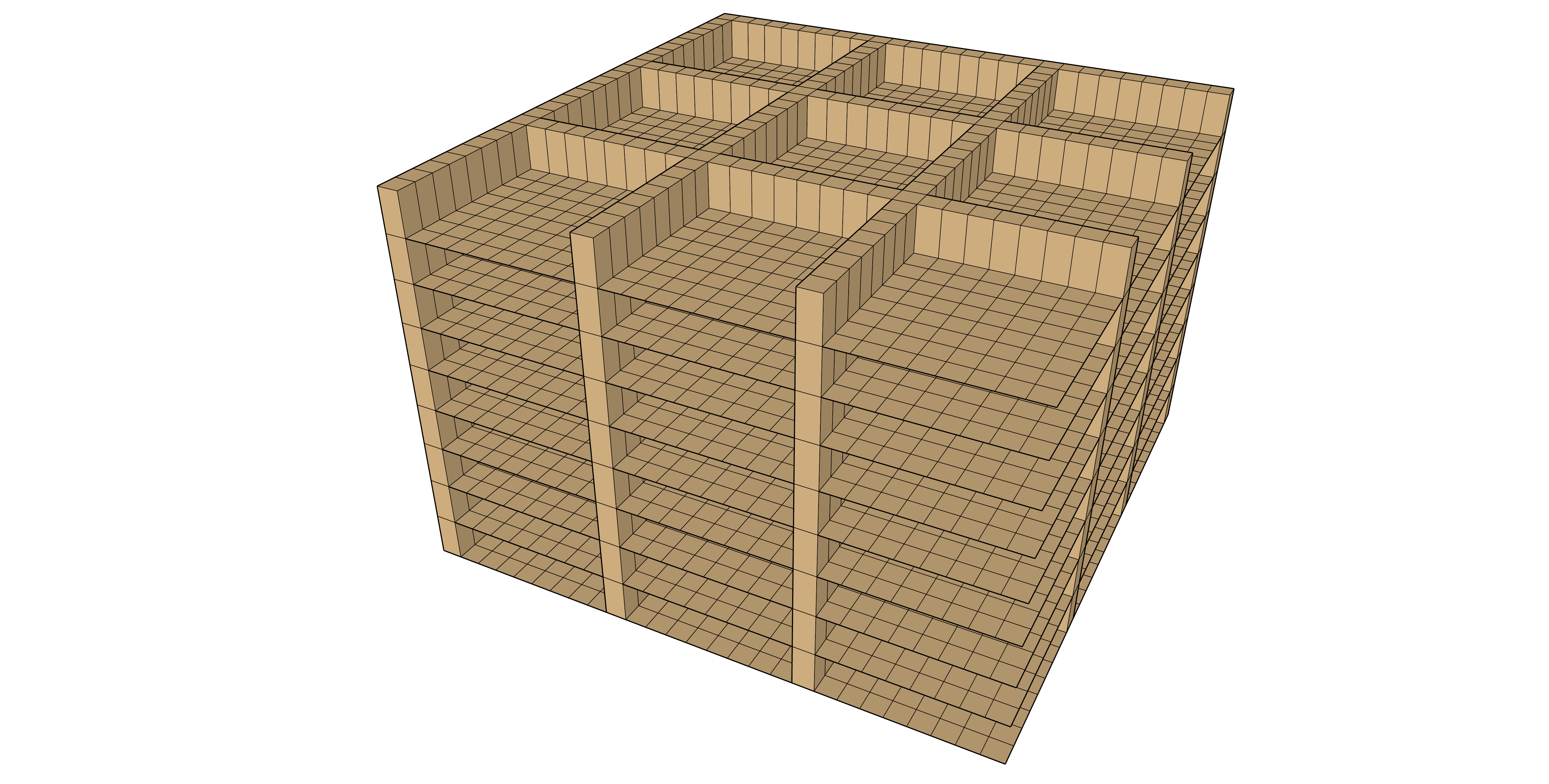}
\caption{Aligned ancilla blocks}\label{fig:decoder1}
\end{subfigure}
\begin{subfigure}{0.48\textwidth}
\includegraphics[width=\linewidth]{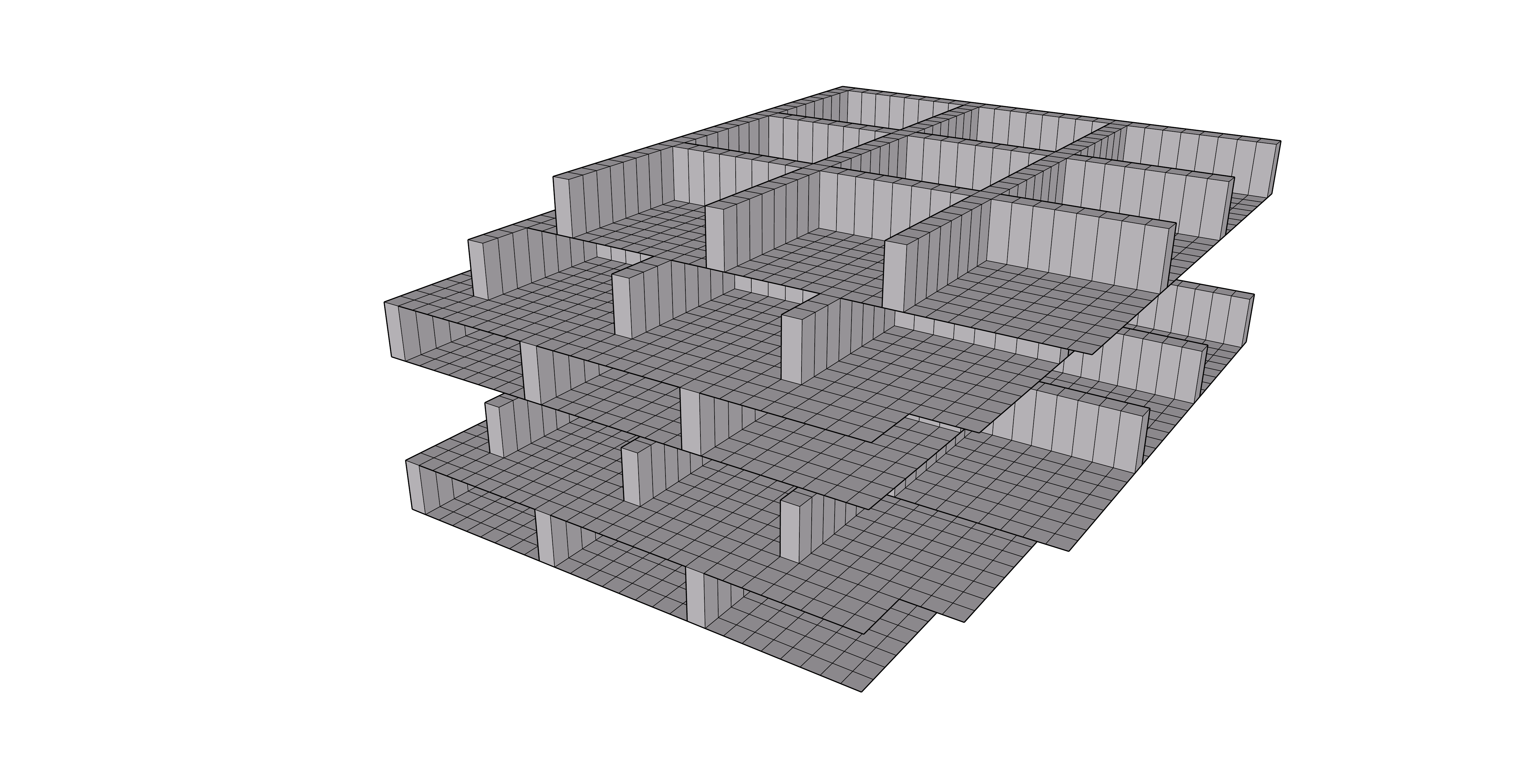}
\caption{Offset ancilla blocks}\label{fig:decoder2}
\end{subfigure}
\caption{
The decoder graphs of the toric code: 
the syndrome bits are vertices; 
the data errors are horizontal edges; and the type-I measurement errors are vertical edges.
The ancilla blocks when aligned leads to time-like correlations between directed repeated measurements. By offsetting the ancilla blocks, the time-like correlations require space-like errors in order to correlate defects from top to bottom.
}
\end{figure*}

\begin{Theorem}\label{thm:4}
In the $i$-th round of error correction, if the correction $\psi'$ creates a \textbf{propagating error}, i.e., $\Pi\psi + \psi'$ contains a path from the time slice $t_i$ to the time slice $t_{i+1}$, then $|\psi[t_i,t_{i+1}]| \ge \frac{1}{2}\left(d(t_i, t_{i+1}) - L\right)$.
\end{Theorem}
\begin{proof}
Suppose $\Pi \psi + \psi'$ contains a path $P_1$ from a vertex $(f_1, t_i)$ to some vertex $(f_2, t_{i+1})$, where $f_1, f_2 \in F$.
$\Pi \psi + \psi'$ must contain another path $P_2$ from a vertex $(f_1', t_i)$ to a vertex $(f_2', t_{i+1})$, where $f_1',f_2' \in F$ and $P_1 \cap P_2 = \emptyset$. We must have 
\begin{equation}\label{eq:1}
    |P_i| = |\Pi\psi \cap P_i| + |\psi' \cap P_i|.
\end{equation}
Let $P_3$ be the shortest path from $(f_1,t_i)$ to $(f_1',t_i)$, $P_4$ be the shortest path from $(f_2,t_{i+1})$ to $(f_2',t_{i+1})$. We must have $|P_3|, |P_4| \le L$. Consider the cycle $C = P_1+P_2+P_3+P_4$.
By the definition of MWPM decoder, $|\psi'| \le |\psi' + C|$, which is equivalent to 
\begin{equation}\label{eq:2}
    |\psi' \cap C| \le |C| - |\psi' \cap C|.
\end{equation}
Combining $|\psi' \cap C| \ge |\psi' \cap P_1| + |\psi' \cap P_2|$ and (\ref{eq:1}), (\ref{eq:2}), we have
\begin{eqnarray}
& &|\psi' \cap P_1| +  |\psi' \cap P_2|\nonumber\\
&\le& |\Pi\psi \cap P_1| +  |\Pi\psi \cap P_2| + |P_3|+|P_4|\nonumber\\
&\le& |\Pi\psi \cap P_1| +  |\Pi\psi \cap P_2| + 2L. \label{eq:3}
\end{eqnarray}
Adding $|\Pi\psi \cap P_1| + |\Pi\psi \cap P_2|$ to both side of (\ref{eq:3}) and combining (\ref{eq:1}), we obtain
\begin{equation}
    2\left(|\Pi\psi \cap P_1| + |\Pi\psi \cap P_2|+L\right) \ge |P_1|+|P_2|.
\end{equation}
Therefore,
\begin{eqnarray}
|\Pi\psi[t_i,t_{i+1}]| &\ge& |\Pi\psi \cap P_1| +  |\Pi\psi \cap P_2|\nonumber\\
&\ge& \frac{1}{2}\left(|P_1|+|P_2| - 2L\right)\nonumber\\
&\ge& d(t_i,t_{i+1}) - L.
\end{eqnarray}
Finally, as $|\Pi \psi [t_i,t_{i+1}]| \le 2 |\psi[t_i,t_{i+1}]|$, we have
\begin{equation}
   |\psi[t_i,t_{i+1}]| \ge \frac{1}{2}\left(d(t_i,t_{i+1}) - L\right).
\end{equation}
\end{proof}

Given the gadgets $\{\gadget_t\}_{t\in \mathbb{N}}$,
we can choose the time slices $\{t_i\}_{i\in\mathbb{N}}$ such that
$d(t_i,t_{i+1}) \ge\alpha L = O(L)$ for some
constant $\alpha \gg 1$. By Theorem~\ref{thm:4}, to make a propagating error happen, the system should have at least $\lceil \frac{(\alpha-1)L}{2} \rceil$ errors.
As $\lceil\frac{L}{2}\rceil \ll \lceil\frac{(\alpha-1)L}{2} \rceil$ errors can already lead to a logical error, assuming the errors are independent, 
the probability of a propagating error is negligible compared to that of a logical error. 

As an extreme case, if we use the Steane-style gadget for all the time, the decoder graph will not have any time-like edges, 
as there are no type-I errors. 
The time slices are not connected with each other so that $d(t,t') = \infty$ for any $t \ne t'$. 
This allows us to choose $t_i = i$. Moreover, when the MWPM decoder matches the defects in time slice $t$, the syndrome information later than $t$ is not used at all. Our analysis is compatible with the fact that Steane's scheme supports single-shot error correction~\cite{Steane:2003}. 

Without taking the complexity of ancilla-state preparation into account, the Steane-style gadget is the best among all transversal gadgets. In practice, however, we might hope that the ancilla block size is bounded by some constant or some function of $L$. 
For example, we would like to partition the toric lattice into $m\times m$ square lattices for some $m$
(for simplicity, we assume that $m$ is a divisor of $L$) 
so that each ancilla block has $2m(m+1) = O(m^2)$ qubits. 
Note that the special case $m = 1$ corresponds to the Shor-style gadget. For general $m$, we naturally expect that the time overhead of fault-tolerant error correction, characterized by $|t_{i+1}-t_i|$, is somewhere between $O(L)$ and $O(1)$.
Unfortunately, if the partition is identical for each time slice, i.e., $\gadget_t = (\tilde{E}, F, \Gamma, \tilde{\partial})$ for all $t \in \mathbb{N}$, 
as depicted in FIG.~\ref{fig:decoder1}, 
we will still have $d(t,t') = |t-t'|$ due to the 
existence of a path of time-like errors $\{(\tilde{e},t)\}_{t\in \mathbb{N}}$, where $\tilde{e} \in \tilde{\partial} F$ is some split edge.
To avoid creating a straight line parallel to the time axis, we need to shuffle the type-I errors by shifting the partitions periodically.

We construct a fault-tolerant error correction scheme as follows. 
For simplicity, we assume that $m = 3k$ for some $k\in \mathbb{N}$. 
We label the rows and columns of the toric lattice from $0$ to $L-1$. As the lattice is periodic, row (column) $L-1$ is adjacent to row (column) $0$. The vertex on row $i$ and column $j$ is labeled by $(i,j)$. At time $t \in \mathbb{N}$, we obtain a gadget $\gadget_t$ by partitioning the lattice into square lattices of size $m\times m$, whose top-left corners are 
$$\Big((pm+kt) \textrm{ mod } L, (qm+kt) \textrm{ mod }L\Big),$$ 
where $p,q = 0, \cdots, L/m-1$.
The decoder graph is demonstrated in FIG.~\ref{fig:decoder2}. 
By our construction, we have $\gadget_t = \gadget_{t+3}$.
One can verify that $d(t,t+3) = t+2 = \Omega(m)$ for any $t \in \mathbb{N}$. Therefore it suffices to choose $|t_{i+1} - t_i| = O(L/m)$ to achieve fault tolerance.

\subsection{Numerical Results}

We study the circuit-level performance of our fault-tolerant error correction schemes numerically.
The $X$- and $Z$-syndrome extractions are applied alternatively.
Our error model is parameterized by a single error parameter $p$ and consists of three parts:
\begin{enumerate}
    \item \textbf{Gate errors}: with probability $p$, each two-qubit CNOT gate is followed by a Pauli error drawn uniformly at random from the set $\{I, X, Y, Z\}^{\otimes 2} \setminus \left\{I \otimes I\right\}$.
    \item \textbf{Measurement errors}: with probability $2p/3$, a measurement outcome in either the $Z$- or $X$-basis is flipped.
    \item \textbf{Preparation errors}: 
    Ancilla preparation can lead to correlated errors that need to be removed through verification or syndrome measurement decoding. Here we assume a simple error model where the complicated ancilla blocks are generated perfectly and then each qubit undergoes an independent depolarizing channel with probability $p_1$, which we set to either $p$ or 0. 
\end{enumerate}
We further simplify by ignoring idling errors which enables us to avoid complications due to scheduling.  Comparison of these syndrome extraction methods for practical application would require a detailed multi-parameter error model, a procedure for ancilla generation and verification, and the connectivity constraints of the quantum processor.

We use a weighted variant of the union-find decoder~\cite{delfosse2017almost,PhysRevA.102.012419} to estimate the fault-tolerance thresholds. 
Table~\ref{table1} and~\ref{table2} compares our block extraction schemes and the conventional Shor's, Steane's and bare-ancilla schemes for the two cases $p_1 = p$ and $p_1 = 0$ respectively.
The ancilla block sizes parameterized by $m$ are fixed. 
As $m$ increases, the threshold is approaching the upper bound provided by Steane's method.
Moreover, although we still need $O(L)$ rounds of extractions when $m$ is fixed, offsetting the ancilla blocks yields higher thresholds than aligning them. 

\begin{table}[h!]
\begin{tabular}{|c|c|c|c|c|c|c|c|}
\hline
     $m$ & Shor & Bare & $3$ & $6$ & $9$ & $12$ & Steane\\
     \hline
Offset & \multirow{2}{*}{0.57\%} & \multirow{2}{*}{0.83\%} & 0.68\% & 0.9\% & 1.05\% & 1.15\% & \multirow{2}{*}{2.05\%}\\
\cline{1-1} \cline{4-7}
Aligned & & & 0.65\% & 0.81\% & 0.91\% & 0.97\% & \\
\hline
\end{tabular}
\caption{ Comparison of thresholds for the case $p_1 = p$.
} \label{table1}
\end{table}

\begin{table}[h!]
\begin{tabular}{|c|c|c|c|c|c|c|c|}
\hline
     $m$ & Shor & Bare & $3$ & $6$ & $9$ & $12$ & Steane\\
     \hline
Offset & \multirow{2}{*}{0.91\%} & \multirow{2}{*}{0.86\%} & 1.14\% & 1.5\% & 1.75\% & 1.91\% & \multirow{2}{*}{3.3\%}\\
\cline{1-1} \cline{4-7}
Aligned & & & 1.14\% & 1.42\% & 1.58\% & 1.68\% & \\
\hline
\end{tabular}
\caption{Comparison of thresholds for the case $p_1 = 0$.}\label{table2}
\end{table}

\section{Conclusion}\label{Sec:5}
In this paper, we have constructed a family of syndrome extraction gadgets for CSS codes described by decomposition of parity check matrices. These gadgets allow us to extract stabilizer elements of the same type in parallel.
Notably, our gadget family includes both 
Shor's~\cite{shor1996fault} and
Steane's~\cite{steane1997active} schemes.
We applied these gadgets on the toric code and construct fault-tolerant error correction schemes. 
Remarkably, for a toric code with distance $L$, one can use ancilla blocks with size $O(m^2)$ to achieve fault tolerance in $O(L/m)$ rounds.
By numerical simulation, we found that our error correction schemes could yield higher thresholds under certain circuit-level noise models.
As the data qubits would be kept idling while preparing the ancilla state, we assume that the idling errors are negligibly small, which could be achieved by trapped-ion qubits~\cite{wang2017single} or other qubits with long coherence time. 
In practice, the ancilla preparation errors will depend on the preparation protocols.
For example, 
as the ancilla blocks inherit the toric code structure, 
one can use the bare-ancilla extractions with post-selections to prepare them~\cite{fowler2012surface}. The ancilla blocks might have correlated errors, but they can be characterized by space-like diagonal edges.
There also exist protocols for preparing a general CSS stabilizer state by state distillations without leaving any correlated errors~\cite{PhysRevA.97.032331}. 
The resource overhead comparisons among the preparation protocols are beyond the scope of this work.

We note our method of syndrome extraction is quite general and only requires the codes be CSS. 
Assuming zero idling error rate, it has been observed on concatenated codes that post-selections can greatly improve the threshold~\cite{Knill2005nature,reichardt2004improved}. 
We conjecture that our methods will yield better circuit-level thresholds than the Shor-style syndrome extraction methods currently used on large distance codes such as 2-dimensional color codes~\cite{chamberland2018flag, PhysRevLett.97.180501}, 
while keeping the rejection rates of post-selections reasonable. 
However, the decoding problem and the time overhead analysis would be more complicated and code specific. We expect that  the framework for analyzing and optimizing the time overhead of sequential Shor-style extractions ~\cite{delfosse2020beyond,delfosse2020short} will be helpful for time optimizing Steane-like extraction.

\section*{Acknowledgements}
We thank Michael Newman and Rui Chao for helpful discussion. The works was sponsored by the NSF EPiQC Expeditions in Computing (1832377), the Office of the Director of National Intelligence - Intelligence Advanced Research Projects Activity through an Army Research Office contract (W911NF-16-1-0082), and the Army Research Office (W911NF-21-1-0005). 

\bibliography{bibliography}
\end{document}